\documentclass[preprint,showpacs,preprintnumbers,superscriptaddress,
amsmath,showkeys,amssymb,aps,pra]{revtex4}

\usepackage{graphicx}
\usepackage{dcolumn}
\usepackage{bm}
\usepackage{times}

\begin{document}

\title{Quantum interference between nuclear excitation by electron 
capture and radiative recombination \footnote{This work is part of the 
doctoral thesis of Adriana Gagyi-P\'alffy, Giessen (D26), 2006.}}

\author{Adriana P\'alffy}
\email{Adriana-Claudia.Gagyi-Palffy@uni-giessen.de}
\affiliation{Institut f\"ur Theoretische Physik,
Justus-Liebig-Universit\"at Giessen, Heinrich-Buff-Ring 16, 35392
Giessen, Germany}

\author{Zolt\'an Harman}
\email{Harman@mpi-hd.mpg.de}
\affiliation{Max-Planck-Institut f\"ur Kernphysik, Saupfercheckweg 1,
69117 Heidelberg, Germany}

\author{Werner Scheid}
\affiliation{Institut f\"ur Theoretische Physik,
Justus-Liebig-Universit\"at Giessen, Heinrich-Buff-Ring 16, 35392
Giessen, Germany}

\date{\today}

\begin{abstract}

We investigate the quantum interference between the resonant process of 
nuclear excitation by electron capture~(NEEC) followed by the radiative 
decay of the excited nucleus, and radiative recombination~(RR). In order 
to derive the interference cross section, a Feshbach projection operator 
formalism is used. The electromagnetic field is considered by means of 
multipole fields. The nucleus is described by a phenomenological 
collective model and by making use of experimental data. The Fano 
profile parameters as well as the interference cross section for 
electric and magnetic multipole transitions in various heavy ions are 
presented. We discuss the experimental possibility of discerning NEEC 
from the RR background.

\end{abstract}



\maketitle


\section{Introduction}


The process of photo-recombination in highly charged heavy ions has been 
the subject of many theoretical and experimental studies up to today, 
concerning both radiative recombination~(RR) and dielectronic 
recombination~(DR)~(see, e.g., \cite{Wolf,Antonio}) and their 
interference. With the enhanced experimental possibilities and achieved 
precision, the subject of electron recombination into highly charged 
ions has been expanding to include QED  
corrections~\cite{Shabaev}. The effect of interference between 
RR with DR has been theoretically studied (see, e.g., 
Ref.~\cite{Zimmermann}) and experimentally 
concluded~\cite{Antonio,Knapp-int,Spies}.

In Ref.~\cite{Goldanskii} a recombination process that is the nuclear 
analog of DR has been theoretically proposed. Although not yet 
experimentally observed, nuclear excitation by electron capture~(NEEC) 
has been an interesting subject after experimental observations of 
atomic physics processes with regard to the structure of the nucleus 
have been recently reported, such as bound-state internal 
conversion~\cite{Carreyre} and nuclear excitation by electron transition 
(NEET)~\cite{Kishimoto}. In the resonant process of NEEC, a free 
electron is recombined into a bound state of an ion with the 
simultaneous excitation of the nucleus. The excited nucleus can then 
decay radiatively or by internal conversion. Several theoretical studies 
have been made concerning NEEC in plasmas~\cite{Goldanskii,Harston} or 
in solid targets~\cite{Cue,Kimball1,Kimball2}. In~\cite{us} we 
presented relativistically correct theoretical cross sections for NEEC 
followed by the radiative decay of the nuclear excited states for 
highly charged heavy ions.

\begin{figure}
\begin{center}
\includegraphics[width=0.60\textwidth]{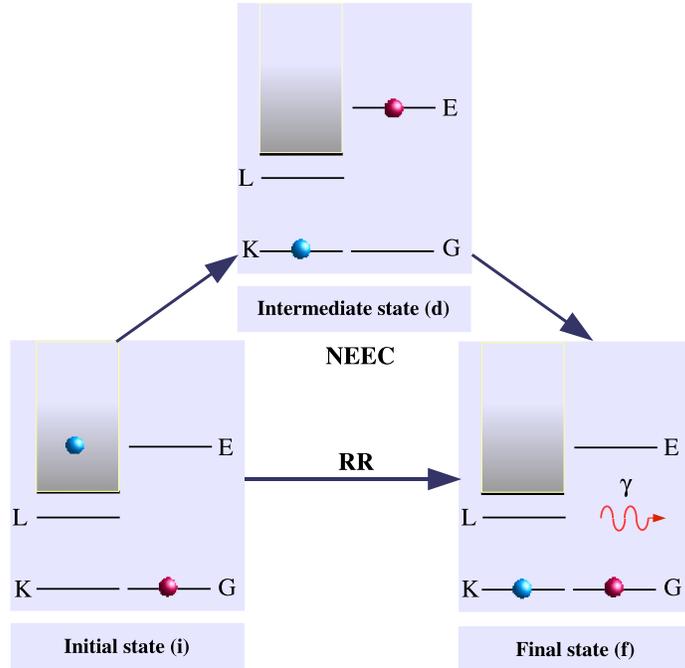}
\caption{\label{schema} NEEC and RR recombination mechanisms of a 
continuum electron into the $K$ shell of a bare ion. The nucleus is 
schematically represented as undergoing the transition from the ground 
state (G) to the excited state (E) and again to its ground state.}
\end{center}
\end{figure}

If the initial and final states for NEEC and RR coincide, quantum 
interference between the two processes occurs. Such an interference 
effect is interesting as it involves two very different pathways: while 
in RR only the recombining electron is involved, NEEC corresponds to a 
quantum path in which the nucleus is excited. In Figure~\ref{schema}  
the RR and NEEC mechanisms are shown schematically. Beside NEEC, the 
strong competing process of RR is always present in an experiment. 
Therefore, the magnitude of the interference effect may also play an 
important role for observing NEEC.

In this paper we theoretically investigate the interference between NEEC 
and RR, focusing on collision systems with suitable excitation energies 
that could be candidates for experimental observation. We derive the 
total cross section of the recombination process with the help of a 
Fesh\-bach projection operator formalism, which allows the separation of 
the interference term from the NEEC and RR cross sections. The radiation 
field is expanded in terms of multipoles in order to clearly discern the 
NEEC transition multipolarities. The electric and 
magnetic electron-nucleus interactions are considered explicitly, and 
the nucleus is described with the help of a geometrical collective model 
and making use of experimental data. The dynamics of the electron is 
governed by the Dirac equation. We express the interference term of the 
cross section using the dimensionless Fano profile parameter for 
electric and magnetic transitions in Sec. II. The numerical results of 
the calculation are given in Sec. III, together with an interpretation 
of the results regarding the possibility of an experimental observation 
of NEEC. We conclude with a short summary. In this work atomic units are 
used unless otherwise specified.


\section{\label{theory} Theoretical formalism for interference effects}


In this section we derive the total cross section of the recombination 
process involving NEEC followed by the radiative decay of the nucleus 
and RR by means of a Feshbach projection operator formalism. We consider 
that the electron is captured into the bound state of a bare ion or an 
ion with a closed-shell configuration. We calculate the interference 
term between NEEC followed by the radiative decay of the excited nucleus 
and RR in the total cross section for electric and magnetic multipole 
transitions of the nucleus.

\subsection{The interference between RR and NEEC in the total cross
section}

The initial state $|\Psi_i\rangle$ of the system describing the nucleus 
in its ground state, the free electron, and the vacuum of the 
electromagnetic field can be written as a direct product of the nuclear, 
electronic, and photonic state vectors:
\begin{equation}\label{eq:initialstate}
| \Psi_i \rangle = | N I_i M_{I_i}, \vec{p} m_s , 0 \rangle \equiv
| N I_i M_{I_i}\rangle \otimes | \vec{p} m_s \rangle \otimes |0 \rangle \,.
\end{equation}
Here, $\vec{p}$ is the asymptotic momentum of the electron, $m_s$ its 
spin projection, and $|N\rangle$ the nuclear ground state, denoted by 
the total angular momentum $I_i$ and its projection $M_{I_i}$. In 
considering RR or NEEC followed by the radiative decay of the nucleus, 
the final state of the recombined system $|\Psi_f\rangle$ consists of 
the nucleus in its ground state, the bound electron and the emitted 
photon. Rather than using the plane wave expansion for the 
electromagnetic field as in~\cite{us}, it is more convenient in this 
case to consider photons of a given angular momentum and parity. The 
final state can be written as
\begin{eqnarray}\label{eq:finalstate}
| \Psi_f \rangle &=& | NI_fM_{I_f} , n_f\kappa_f m_f , \lambda kLM \rangle\\
&\equiv& | N I_fM_{I_f}\rangle \otimes
| n_f\kappa_f m_f \rangle \otimes |\lambda k LM \rangle \ ,\nonumber
\end{eqnarray}
with $n_f$, $\kappa_f$, and $m_f$ being the principal, Dirac angular 
momentum, and magnetic quantum numbers of the bound one-electron state, 
respectively. The emitted photon has the wave number $k$, the total 
angular momentum $L$ and its projection $M$. Furthermore, $\lambda$ 
stands for electric $(e)$ or magnetic $(m)$ waves. The intermediate 
resonant state formed by the electron capture in the process of NEEC 
consists of the excited nucleus, the bound electron, and the vacuum 
state of the electromagnetic field,
\begin{eqnarray}
| \Psi_d \rangle &=& | N^*I_dM_{I_d}, n_d\kappa_d m_d ,0 \rangle \\
&\equiv& | N^*I_dM_{I_d} \rangle \otimes
| n_d\kappa_d m_d \rangle \otimes |0 \rangle \ . \nonumber
\end{eqnarray}
The excited nuclear state is denoted by $|N^*\rangle$. In our case, the 
recombined electron does not undergo further decay cascades, i.e., 
$n_d=n_f$, $\kappa_d=\kappa_f$ and $m_d=m_f$.

Following the formalism presented in Ref.~\cite{us}, we introduce 
projector operators onto the individual subspaces, in order to separate 
these states in the perturbative expansion of the transition operator. 
We neglect corrections due to two- or more-photon states \cite{Zakowicz,Zakowicz2} 
and due to the presence of the negative electronic continuum. The Fock 
space is then given by the sum of three subspaces: the subspace of the 
states that contain the free electron, with its projector operator $P$, 
the subspace of the states characterized by the presence of the excited 
nucleus, together with the corresponding projector operator $Q$, and 
finally the subspace of the states with a photon, associated to the 
projector operator $R$. We postulate the completeness relation
\begin{equation}
P+Q+R={\bf 1}\ ,
\end{equation}
where ${\bf 1}$ is the unity operator of the total Fock space.

The total Hamiltonian operator for the system consisting of the nucleus 
($n$), the electron ($e$), and the radiation field ($r$) can be written 
as
\begin{equation}\label{eq:totalh}
H = H_n + H_e + H_r + H_{en} + H_{er} + H_{nr} \,.
\end{equation}
The expressions of the first three Hamiltonians can be found 
in~\cite{us}. Interactions between the three subsystems are described by 
the three remaining Hamiltonians in Eq.~(\ref{eq:totalh}). We adopt the 
Coulomb gauge for the electron-nucleus interaction ($en$) because it 
allows the separation of the dominant Coulomb attraction between the 
electronic and the nuclear degrees of freedom:
\begin{equation}\label{eq:coulomb}
H_{en}=\int d^3r_n\frac{\rho_n(\vec{r}_n)}{|\vec{r}_e-\vec{r}_n|}\ .
\end{equation} 
Here, $\rho_n(\vec{r}_n)$ is the nuclear charge density and the 
integration is performed over the whole nuclear volume. The static part 
of the electron-nucleus interaction is contained in Hamiltonian $H_e$. 
The interaction of the electron with the transverse photon field 
quantized in the volume of a sphere of radius $R$ is given by
\begin{equation}\label{eq:radhamilton}
H_{er} =-\vec{\alpha}\cdot\vec{A}=-\sum_{\lambda k LM} \left(a^{\dagger}_{\lambda kLM}\vec{\alpha}\cdot\vec{A}_{\lambda k LM}(\vec{r})+{\rm H.c.} \right)\ ,
\end{equation}
with the vector potential of the quantized electromagnetic 
field~\cite{Ring}
\begin{equation}
\vec{A}(\vec{r})=\sum_{\lambda k LM}\left(\vec{A}_{\lambda k LM}(\vec{r})\,
a^{\dagger}_{\lambda k LM}+\vec{A}^*_{\lambda k LM}(\vec{r})\,a_{\lambda k LM}
\right)\ .
\end{equation} 
Here, $\vec\alpha$ is the vector of the Dirac matrices and the two 
independent solutions of the wave equation for the $\vec{A}_{\lambda k 
LM}(\vec{r})$ are
\begin{eqnarray}\label{wave_sol}
\vec{A}_{(m)kLM}(\vec{r}) &=& \sqrt{\frac{4\pi ck}{R}}j_L(kr)
\vec{Y}^M_{LL}(	\theta,\varphi)\ ,\\
\vec{A}_{(e)kLM}(\vec{r}) &=& \frac{i}{k}\sqrt{\frac{4\pi ck}{R}}\vec{\nabla}\times\big(j_L(kr)
\vec{Y}^M_{LL}(	\theta,\varphi)\big) \ , \nonumber 
\end{eqnarray}
where the quantum number $k$ is discretized by requiring the proper 
boundary conditions at a perfectly conducting sphere of radius $R$. The 
$\vec{Y}^M_{LL}( \theta,\varphi)$ denote the vector spherical harmonics, 
given by \cite{Edmonds}
\begin{equation}
\vec{Y}^M_{LL}(	\theta,\varphi)=\sum_\nu\sum_q C(L\ 1\ L;\nu\ q\ M)Y_{L\nu}(\theta,\varphi)\vec{\epsilon}_q\ ,
\end{equation}
where $q=0,\pm 1$ and the spherical unit vectors $\vec{\epsilon}_q$ 
expressed in terms of the Cartesian unit vectors 
$\left(\vec{e}_x,\vec{e}_y,\vec{e}_z\right)$ are
\begin{eqnarray}
\vec{\epsilon}_+ &=& -\frac{1}{\sqrt{2}}(\vec{e}_x+i\vec{e}_y)\ ,\\
\vec{\epsilon}_0 &=& \vec{e}_z \ ,\nonumber\\ 
\vec{\epsilon}_- &=& \frac{1}{\sqrt{2}}(\vec{e}_x-i\vec{e}_y) \,.\nonumber
\end{eqnarray}

Similarly, the interaction of the nucleus with the electromagnetic field 
is given by the Hamiltonian
\begin{equation}\label{eq:hnr}
H_{nr} = -\frac{1}{c}\sum_{\lambda k LM} \bigg( a^{\dagger}_{\lambda k LM}\int d^3r_n\vec{j}_n(\vec{r}_n)
\cdot\vec{A}_{\lambda k LM}(\vec{r}_n) + {\rm H.c.} \bigg) \ ,
\end{equation}
where  $\vec{j}_n(\vec{r}_n)$ is the nuclear current.

Using the projection operators we can separate the perturbation $V$ in 
the total Hamiltonian
\begin{equation}
H = H_0 + V \ ,
\end{equation}
with
\begin{eqnarray}
H_0 &=& PHP + QHQ + RHR \,,  \label{eq:h_0}\\
V &\equiv& H - H_0 = PHQ + QHP + PHR \\
&+& RHP + RHQ + QHR \,. \nonumber 
\end{eqnarray}
This way of defining $H_0$ has the advantage that the effect of the 
nuclear potential on bound and continuum electron states is included in 
$H_0$ to all orders. The individual terms in the perturbation operator 
describe transitions between the different subspaces. For example, $QHP$ 
describes in the lowest order the time-reversed process of internal 
conversion (IC), namely, NEEC, while $PHR$ and $RHP$ are the first-order 
operators for photo-ionization and radiative recombination, respectively.

The transition operator is defined as~\cite{Taylor}
\begin{equation}
T(z) = V + VG(z)V \,,
\end{equation}
where  $G(z)$ is the Green operator of the total system given by
\begin{equation}
G(z)= (z-H)^{-1} \,.
\end{equation}
Here, $z$ is a complex energy variable. The total cross section for a 
process can be expressed by the modulus square of the matrix element of 
the transition operator, after summing over the final states and 
averaging over the initial states that are not resolved in the 
experiment,
\begin{eqnarray}\label{eq:tsigma}
\sigma_{i \to f}(E) &=& \frac{2\pi}{F_i}\sum_{M_{I_f}m_d}
\sum_{\lambda LM} \frac{1}{2(2I_i +1)} \sum_{M_{I_i} m_s} \\
&& \frac{1}{4\pi} \int d\Omega_p
\lim_{\epsilon \to 0+}  |\langle \Psi_f|T(E+i\epsilon) |\Psi_i\rangle|^2 \rho_f \,, \nonumber
\end{eqnarray}
with the $\Psi_f$ and $\Psi_i$ as final and initial eigenstates of 
$H_0$, respectively [see Eqs.~(\ref{eq:initialstate}) and 
(\ref{eq:finalstate})]. Here, $F_i$ denotes the flux of the incoming 
electrons, $\rho_f$ the density of the final photonic states, and 
$\Omega_p$ is the direction of the incoming free electron characterized 
by the angles $\theta_p$ and $\varphi_p$.

We use the Lippmann-Schwinger equation 
\begin{eqnarray}
G(z) &=& G_0(z) + G_0(z)VG_0(z) \\ 
     &+& G_0(z)VG_0(z)VG_0(z) + \dots \nonumber
\end{eqnarray}	
to write the perturbation series for $T(z)$ in powers of $V$ with the 
Green function $G_0(z)$ of the unperturbed Hamiltonian $H_0$:
\begin{equation}
T(z) = V + VG_0(z)V + VG_0(z)VG_0(z)V + \dots \,.
\end{equation}
Since the initial state of the NEEC process is by definition an 
eigenstate of $P$, and the final state is an eigenstate of $R$, we only 
need to consider the projection $RTP$ of the transition operator:
\begin{eqnarray}\label{eq:rtp}
RT(z)P &=& RVP + RVG_0(z)VP \\
&+& RVG_0(z)VG_0(z)VP \nonumber \\
&+& RVG_0(z)VG_0(z)VG_0(z)VP + \dots  \,. \nonumber
\label{exp}
\end{eqnarray}
The first term $RVP$ accounts for the radiative recombination. Taking 
into account from the infinite perturbation expansion in 
Eq.~(\ref{eq:rtp}) the terms that correspond to NEEC \cite{us} we can 
write the final expression for the transition amplitude for the 
recombination process as
\begin{eqnarray}
\langle \Psi_f |RT(z)P|\Psi_i \rangle &=&
\langle\Psi_f |RH_{er}P|\Psi_i\rangle \\
&+&\sum_{d}\frac{\langle \Psi_f | H_{nr} |\Psi_d \rangle
\langle \Psi_d| H_{en}+H_{magn} |\Psi_i \rangle}
{z-E_d + \frac{i}{2} \Gamma_d} \,. \nonumber
\end{eqnarray}
Here, $\Gamma_d$ denotes the total natural width of the excited state 
$|d\rangle=|N^*I_d M_{I_d},n_d\kappa_d m_d, 0\rangle$. The magnetic 
interaction Hamiltonian $H_{magn}$ accounts for the recombination of the 
free electron by exchanging a virtual transverse photon with the nucleus 
in the unretarded approximation~\cite{us},
\begin{equation}
H_{magn} = - \frac{1}{c} \vec{\alpha} \int d^3r_n 
\frac{\vec{j}_n(\vec{r}_n)}{|\vec{r}-\vec{r}_n|}\ .
\label{hmagn}
\end{equation}
Using the expression of the transition operator, the total cross section 
can then be written as
\begin{eqnarray}
\sigma_{i \to f}(E) &=& \frac{2\pi}{F_i}\sum_{M_{I_f}m_d} \sum_{\lambda LM} \frac{1}{2(2I_i +1)}
\sum_{M_{I_i} m_s} \frac{1}{4\pi} \int d\Omega_p \\*
&&\Big|\langle NI_fM_{I_f},n_d\kappa_d m_d,\lambda kLM|
H_{er}|NI_iM_{I_i},\vec{p}m_s,0 \rangle \nonumber \\*
&&+ \sum_{M_{I_d}}\frac{\langle N I_f M_{I_f}, n_d\kappa_d m_d , \lambda k LM | H_{nr} 
|N^* I_d M_{I_d},n_d\kappa_d m_d,0\rangle} {(E-E_d) + \frac{i}{2} \Gamma_d} \nonumber \\*
&&\times \langle N^* I_d M_{I_d},n_d\kappa_d m_d, 0| H_{en}+H_{magn} 
| N I_i M_{I_i} , \vec{p} m_s , 0 \rangle \Big|^2 \rho_f \ . \nonumber
\end{eqnarray}
The first term in the modulus squared accounts for RR and the second one 
for NEEC. We can separate therefore the equation above in three terms,
\begin{equation}
\sigma_{i \to f}(E)=\sigma_{\rm RR}(E)+\sigma_{\rm NEEC}(E)+\sigma_{\rm int}(E)\ ,
\label{tcs}
\end{equation}
with the RR and NEEC total cross sections given by
\begin{eqnarray}
\sigma_{\rm RR}(E)&=&\frac{2\pi}{F_i}\sum_{M_{I_f}m_d} \sum_{\lambda LM} \frac{1}{2(2I_i +1)} \sum_{M_{I_i} m_s} \\*
&&\frac{1}{4\pi} \int d\Omega_p 
|\langle NI_fM_{I_f},n_d\kappa_d m_d,\lambda kLM|H_{er}|NI_iM_{I_i},\vec{p}m_s,0 \rangle|^2\rho_f, \nonumber
\end{eqnarray}
and
\begin{eqnarray}
\sigma_{\rm NEEC}(E)&=&\frac{2\pi}{F_i}\sum_{M_{I_f}m_d} \sum_{M_{I_d}}\sum_{\lambda LM} \frac{1}{2(2I_i +1)} \sum_{M_{I_i} m_s}\\*
&&\frac{1}{4\pi} \int d\Omega_p 
\bigg|\frac{\langle N I_f M_{I_f}, n_d\kappa_d m_d , \lambda k LM | H_{nr} |N^* I_d M_{I_d},n_d\kappa_d m_d,0\rangle}
{(E-E_d) + \frac{i}{2} \Gamma_d} \nonumber \\*
&&\times \langle N^* I_d M_{I_d},n_d\kappa_d m_d, 0| H_{en}+H_{magn} 
| N I_i M_{I_i} , \vec{p} m_s , 0 \rangle \bigg|^2 \rho_f \,. \nonumber
\end{eqnarray}
The term describing the interference between RR and NEEC can be written as
\begin{eqnarray}\label{sint}
\sigma_{\rm int}(E)&=&\frac{2\pi}{F_i}\sum_{M_{I_f}m_d} \sum_{M_{I_d}}\sum_{\lambda LM} \frac{\rho_f}{2(2I_i +1)} \sum_{M_{I_i} m_s} \\*
&&\frac{1}{4\pi} \int d\Omega_p 
\bigg(\frac{\langle N I_f M_{I_f}, n_d\kappa_d m_d , \lambda k LM | H_{nr}
|N^* I_d M_{I_d},n_d\kappa_d m_d,0\rangle}
{(E-E_d) + \frac{i}{2} \Gamma_d} \nonumber \\*
&&\times \langle N^* I_d M_{I_d},n_d\kappa_d m_d, 0| H_{en}+H_{magn}
| N I_i M_{I_i} , \vec{p} m_s , 0 \rangle \nonumber \\*
&&\times \langle NI_fM_{I_f},n_d\kappa_d m_d,\lambda kLM|H_{er}
|NI_iM_{I_i},\vec{p}m_s,0 \rangle^* +{\rm H.c.}\bigg)\,. \nonumber
\end{eqnarray}

The aim of this paper is to calculate the interference term in the total 
cross section. The calculation of the NEEC cross section and predicted 
values for several collisions systems can be found in~\cite{us}. 
Furthermore, the calculation of the RR total cross section is well 
understood. An extensive tabulation of relativistic total cross sections 
for RR as a function of energy ranging from closely above the threshold 
to the relativistic regime of relative electron energies is available 
in~\cite{Eich}.

If we consider the matrix element of the Hamiltonian $H_{er}$ connecting 
the radiation field and the electrons in the interference term, the 
initial and the final nuclear total angular momenta as well as their 
projections have to coincide, as they are not influenced by RR,
\begin{eqnarray}
&&\langle NI_fM_{I_f},n_d\kappa_d m_d,\lambda kLM|H_{er}
|NI_iM_{I_i},\vec{p}m_s,0 \rangle \quad \quad \\
&&=\delta_{I_iI_f} \delta_{M_{I_f}M_{I_i}}\langle n_d\kappa_d m_d,\lambda kLM|
H_{er}|\vec{p}m_s,0 \rangle\,. \nonumber
\end{eqnarray}

The initial state continuum electronic wave function is given through 
the partial wave expansion~\cite{Eichler}
\begin{equation}
|\vec{p} m_s\rangle=\sum_{\kappa m m_l}i^l e^{i\Delta_{\kappa}} Y_{l m_l}^*(\Omega_p)
C\left(l\ \frac{1}{2}\ j;m_l\ m_s \ m\right)| \varepsilon\kappa m\rangle\ ,
\end{equation}
where $\varepsilon$ is the energy of the continuum electron measured 
from the ionization threshold, $\varepsilon=\sqrt{p^2c^2+c^4}-c^2$. The 
orbital angular momentum of the partial wave is denoted by $l$ and the 
corresponding magnetic quantum number by $m_l$, while the partial wave 
phases $\Delta_{\kappa}$ are chosen so that the continuum wave function 
fulfills the boundary conditions of an incoming plane wave and an 
outgoing spherical wave. The total angular momentum quantum number of 
the partial wave is $j=|\kappa|-\frac{1}{2}$. The interference cross 
section in the case of NEEC involving a nuclear transition with specific 
parity $\lambda$ and multipolarity $L$ can then be written as
\begin{eqnarray}
\sigma_{\rm int}&=&\frac{2\pi}{F_i}\sum_{M_{I_d}M_{I_i}} \sum_{Mm_d}
\frac{\rho_f}{2(2I_i +1)} \sum_{\kappa m } \frac{1}{4\pi} \\*
&\times&\bigg(\frac{\langle N I_i M_{I_i},  \lambda k LM | H_{nr}
|N^* I_d M_{I_d},0\rangle} {(E-E_d) + \frac{i}{2} \Gamma_d}
\langle N^* I_d M_{I_d},n_d\kappa_d m_d| H_{en}+H_{magn}
| N I_i M_{I_i} ,\varepsilon\kappa m   \rangle\nonumber \\*
&\times& \langle n_d\kappa_d m_d,\lambda kLM|H_{er}
|\varepsilon\kappa m,0 \rangle^*+{\rm H.c.}\bigg)\,. \nonumber
\end{eqnarray}
We can relate  the interference cross section term with  the NEEC
cross section, introducing the dimensionless Fano profile parameter
$Q_f$. The expression of the  NEEC cross section  from Ref.~\cite{us} is
\begin{equation}
\sigma_{i \to d \to f}(E) = \frac{2\pi^2}{p^2}
\frac{A_r^{d \to f} Y_n^{i \to d}}{\Gamma_d} L_d(E-E_d) \ ,
\end{equation} 
where $A_r^{d \to f}$ is the radiative rate defined as
\begin{equation}
A_r^{d \to f} = \frac{2\pi}{2I_d + 1} \sum_{M_{I_f} M} \sum_{M_{I_d}}
|\langle N I_f M_{I_f}, n_d\kappa_d m_d , \lambda kLM| H_{nr}
| N^* I_d M_{I_d},n_d\kappa_d m_d, 0\rangle |^2\rho_f 
\end{equation}
and $Y_n^{i \to d}$ is the NEEC rate,
\begin{equation}
Y_n^{i \to d}=\frac{2\pi}{2(2I_i+1)}\sum_{{M_{I_i}} m_s}\sum_{M_{I_d} m_d}
\int d\Omega_p
|\langle N^*I_dM_{I_d},n_d\kappa_d m_d,0|H_{en}+H_{magn}
|NI_iM_{I_i},\vec{p}m_s,0\rangle|^2\rho_i\,. \label{Yrate} 
\end{equation}
Furthermore, $p$ denotes the continuum electron momentum and $\rho_i$ 
the density of the initial electronic states. The explicit energy 
dependence of the interference term can be expressed with the help of 
the Lorentz profile $L_d(E-E_d)$, defined as
\begin{equation}
L_d(E-E_d) = \frac{\Gamma_d / 2\pi}{(E-E_d)^2 + \frac{1}{4} \Gamma_d^2}\ ,
\label{lorentz}
\end{equation}
which in turn is related to the NEEC total cross section. The 
interference cross section can be written in the concise form 
\cite{Zimmermann}
\begin{equation}
\sigma_{\rm int}=\sigma_{\rm NEEC}\frac{\Gamma_d}{Y_n^{i\to d}}\frac{2I_d +1}{2I_i +1}
\left(
2\frac{E-E_d}{\Gamma_d}{\rm Re}\left(\frac{1}{Q_f}\right)
+{\rm Im}\left(\frac{1}{Q_f}\right)\right)\ ,
\end{equation}
with the dimensionless Fano profile parameter
\begin{eqnarray}\label{fano}
\frac{1}{Q_f}&=&\pi\rho_i\sum_{M_{I_d}M_{I_i}} \sum_{Mm_d}  \sum_{\kappa m }
\langle N^* I_d M_{I_d},n_d\kappa_d m_d| H_{en}+H_{magn}
| N I_i M_{I_i} ,\varepsilon\kappa m \rangle \\* 
&\times& \frac{\langle N I_i M_{I_i},  \lambda k LM | H_{nr}
|N^* I_d M_{I_d},0\rangle\langle n_d\kappa_d m_d,\lambda kLM|H_{er}
|\varepsilon\kappa m,0 \rangle^*}{\sum_{M'_{I_i} M'} \sum_{M'_{I_d}}
\big|\langle N I_i M'_{I_i} , k\lambda LM'| H_{nr}
|N^* I_d M'_{I_d}, 0\rangle \big|^2}\,.\nonumber
\end{eqnarray}
We have used prime indices for the summations in the expression of the 
nuclear radiative rate in the denominator in order to avoid
confusion. With the further observation that the Fano profile parameter 
$1/Q_f$ is real for both the electric and magnetic cases, the 
interference cross section yields
\begin{equation}
\sigma_{\rm int}=\sigma_{\rm NEEC}\frac{2(E-E_d)}{Y_n^{i \to d}}
\frac{2I_d +1}{2I_i +1}\frac{1}{Q_f}\,.
\end{equation}

\subsection{Electric transitions}

In order to calculate the matrix elements in the Fano profile parameter 
in Eq.~(\ref{fano}), an adequate nuclear model is needed. Following the 
outline in~\cite{us}, we describe the nucleus by means of a geometrical 
collective model~\cite{Greiner} which assumes that the excitations of 
the nucleus are vibrations and rotations of the nuclear surface, which 
is parameterized as
\begin{equation}
R(\theta,\varphi,t)=R_0\Big(1+\sum_{\ell=0}^\infty\sum_{m=-\ell}^{\ell} 
\alpha_{\ell m}^*(t)Y_{\ell m}(\theta,\varphi)\Big)\ .
\label{param}
\end{equation}
The time-dependent deformation amplitudes $\alpha_{\ell m}(t)$ describe 
the nuclear surface with respect to a sphere of radius $R_0$ and serve 
as collective coordinates.  This parameterization can be used to 
calculate the matrix element corresponding to the NEEC process for a 
given partial wave component and a given multipolarity $L$, that yields 
\cite{us}
\begin{eqnarray}\label{elel}
&&\langle N^*I_dM_{I_d}, n_d\kappa_dm_d|H_{en}
|NI_i M_{I_i},\varepsilon\kappa m \rangle \\*
&=&\sum_{\mu=-L}^L(-1)^{I_d+M_{I_i}+L+\mu+m+3j_d}
R_0^{-(L+2)} R_{L,\kappa_d,\kappa}
\langle N^* I_d\|Q_L\|NI_i\rangle \nonumber \\*
&\times&\sqrt{2j_d+1}\sqrt{\frac{4\pi}{(2L+1)^3}}
C(I_i\ I_d\ L;-M_{I_i}\ M_{I_d} \ \mu) C(j\ j_d\ L;-m\ m_d\ -\mu)
C\left(j_d\ L\ j;\frac{1}{2}\ 0 \ \frac{1}{2}\right)\,, \nonumber
\end{eqnarray}
where $Q_{LM}$ is the electric multipole moment defined by \cite{Ring}
\begin{equation}
Q_{LM}=\int d^3r_n r_n^{L} Y_{LM}(\theta_n,\varphi_n)\rho_n(\vec{r}_n)\ .
\end{equation}
The electronic radial integral is given by
\begin{eqnarray}\label{rrs}
R_{L,\kappa_d,\kappa} &=& \frac{1}{R_0^{L-1}}\int_0^{R_0} dr r^{L+2}
\left(f_{n_d \kappa_d}(r)f_{\varepsilon\kappa}(r)+
g_{n_d \kappa_d}(r)g_{\varepsilon\kappa}(r)\right)+ \\*
&+&R_0^{L+2}\int_{R_0}^\infty dr r^{-L+1}
\left(f_{n_d \kappa_d}(r)f_{\varepsilon\kappa}(r)
+g_{n_d \kappa_d}(r)g_{\varepsilon\kappa}(r)\right)\,. \nonumber
\end{eqnarray}
with $g_{\varepsilon\kappa}(r)$ and
$f_{\varepsilon\kappa}(r)$ being the large and small radial components of
the relativistic continuum electron partial wave function
\begin{equation}
\Psi_{\varepsilon\kappa m}(\vec{r})=\left(\begin{array}{c} 
g_{\varepsilon\kappa}(r)\Omega_{\kappa}^{m}(\theta_e,\varphi_e)\\if_{\varepsilon\kappa}(r)\Omega_{-\kappa}^{m}(\theta_e,\varphi_e)\end{array} 
\right)\,,
\end{equation}
with the spherical spinor functions $\Omega_{\kappa}^{m}$, and $g_{n_d 
\kappa_d}(r)$ and $f_{n_d \kappa_d}(r)$ the radial components of the 
bound Dirac wave function.

For the matrix element of the interaction Hamiltonian~(\ref{eq:hnr}) 
between the nucleus and the radiation field, we follow the outline 
in~\cite{Greiner_E}, considering that the wavelength of the radiation is 
large compared to the nuclear radius, $kR_0\ll 1$, so that the Bessel 
functions can be approximated in the first order in $kr$ as
\begin{equation}
j_L(kr)\simeq \frac{(kr)^L}{(2L+1)!!}\ .
\label{lwl}
\end{equation}
In this case the electric solution of the wave equation can be written as
\begin{equation}
\vec{A}_{(e)kLM}(\vec{r})= -\sqrt{\frac{4\pi ck}
{R}}\frac{\sqrt{(L+1)(2L+1)}}{(2L+1)!!}(kr)^{L-1}\vec{Y}^M_{LL-1}(\theta,\varphi)
\ .
\end{equation}
With the use of the continuity equation for the nuclear current 
$\vec{j}_n$ we obtain for the matrix element
\begin{eqnarray}\label{elnr}
\langle N I_i M_{I_i},(e)kLM |H_{nr}|N^* I_d M_{I_d},0\rangle
&=&(-1)^{I_d-M_{I_d}+1}\sqrt{\frac{4\pi ck}{R}}
C\left(I_i\ I_d\ L; M_{I_i}\ -M_{I_d}\ M\right) \quad \\*
&\times& \frac{\sqrt{L+1}}{\sqrt{L(2L+1)}}\frac{ik^L}{(2L+1)!!}
\langle N I_i  \|Q_L\|N^* I_d \rangle \,. \nonumber
\end{eqnarray}
The remaining matrix element of $H_{er}$ can be evaluated by writing the 
electric solution of the wave equation in Eq.~(\ref{wave_sol}) in a more 
suitable form. Using the properties of the vector spherical harmonics 
\cite{Varshalovich} we obtain
\begin{eqnarray}
\vec{A}_{(e)kLM}(\vec{r})&=&\sqrt{\frac{4\pi ck}{R}}\left(
\sqrt{\frac{L}{2L+1}}j_{L+1}(kr)\vec{Y}^M_{LL+1}(\theta,\varphi)\right. \\*
&-&\left.\sqrt{\frac{L+1}{2L+1}}j_{L-1}(kr)\vec{Y}^M_{LL-1}
(\theta,\varphi)\right)\,.  \nonumber
\end{eqnarray}
The electron-radiation interaction matrix element then yields
\begin{eqnarray}
&&\langle n_d\kappa_d m_d,(e) kLM|H_{er}|\varepsilon\kappa m,0\rangle \\*
&=&-\sqrt{\frac{4\pi ck}{R}} \bigg(
\sqrt{\frac{L}{2L+1}}
\langle n_d\kappa_d m_d|j_{L+1}(kr) 
\vec{\alpha}\cdot\vec{Y}^M_{LL+1}(\theta,\varphi) |\varepsilon\kappa m\rangle
\nonumber \\*
&-&\sqrt{\frac{L+1}{2L+1}}\langle n_d \kappa_d m_d|
j_{L-1}(kr)\vec{\alpha}\cdot\vec{Y}^M_{LL-1}(\theta,\varphi)
|\varepsilon\kappa m\rangle\bigg)\ . \nonumber
\label{Her_me}
\end{eqnarray}
The matrix elements containing the product of the Bessel spherical 
functions, the Dirac matrix $\vec{\alpha}$ and the vector spherical 
harmonics can be expressed in a compact way using the properties of the 
spherical tensor operators~\cite{Grant}. The expression in the above 
equation becomes
\begin{eqnarray}\label{eler}
&&\langle n_d\kappa_d m_d,(e) kLM|H_{er}|\varepsilon\kappa m,0\rangle\\*
&&=i(-1)^{j-L+\frac{1}{2}}\sqrt{\frac{4\pi ck}{R}}C(j\ L\ j_d;m\ M\ m_d)
\sqrt{\frac{2j+1}{4\pi}}
\left(\begin{array}{ccc}j_d & j& L\\\frac{1}{2}&-\frac{1}{2}&0\end{array}\right)
\nonumber \\*
&&\times \bigg[\sqrt{\frac{L+1}{L(2L+1)}}(LI^-_{L-1}-(\kappa_d-\kappa)I^+_{L-1})
\nonumber \\
&&
\ \ \ \ \  +  \sqrt{\frac{L}{(L+1)(2l+1)}}
((L+1)I^-_{L+1}+(\kappa_d-\kappa)I^+_{L+1})\bigg]\ , \nonumber
\end{eqnarray}
with the radial integrals
\begin{equation}\label{eq:iint}
I^{\pm}_L=\int_0^{\infty}drr^2j_L(kr)\left(
g_{n_d \kappa_d}(r)f_{\varepsilon\kappa}(r)
\pm g_{\varepsilon\kappa}(r)f_{n_d \kappa_d}(r)\right)\,.
\end{equation}
Combining the formulas of the three matrix elements from 
Eqs.~(\ref{elel}), (\ref{elnr}) and (\ref{eler}) in the expression of 
the Fano profile parameter $Q_f^{(e)}$ and using the summation 
properties of the Clebsch-Gordan coefficients we obtain the final 
formula
\begin{eqnarray}
\frac{1}{Q_f^{(e)}}&=&\pi\rho_i  (-1)^{3I_d+I_i+1}R_0^{-(L+2)} (2j_d+1)
\sqrt{\frac{L}{(L+1)(2L+1)^3}} \\*
&\times& k^{-L}(2L+1)!! \sum_{\kappa  }R_{L,\kappa_d,\kappa}(2j+1)
\left(\begin{array}{ccc}j_d & j& L\\
\frac{1}{2}&-\frac{1}{2}&0\end{array}\right)^2
\nonumber \\
&\times&
\bigg[\sqrt{\frac{L+1}{L(2L+1)}}(LI^-_{L-1}-(\kappa_d-\kappa)I^+_{L-1}) \nonumber \\*
&&+ \sqrt{\frac{L}{(L+1)(2L+1)}}((L+1)I^-_{L+1}+(\kappa_d-\kappa)I^+_{L+1})\bigg]\ .
\label{fanoe} \nonumber
\end{eqnarray}

\subsection{Magnetic transitions}

The magnetic transitions in the nucleus can be easily included in the 
calculation by assuming that the electron does not penetrate the 
nucleus, i.e., that the electronic radial coordinate $r_e>r_n$ is always 
larger than the nuclear radial coordinate. This approximation is precise 
enough for the studied cases~\cite{Rose,Alder}. The NEEC matrix element 
for the magnetic transition, involving only the magnetic Hamiltonian 
$H_{magn}$ for a given partial wave and a given multipolarity can be 
written as~\cite{us}
\begin{eqnarray}\label{hmg}
&&\langle N^*I_dM_{I_d},n_d \kappa_dm_d|H_{magn}|N I_i M_{I_i},\varepsilon \kappa m\rangle= \\* 
&&4\pi i\sqrt{\frac{L+1}{L(2L+1)^3}}
\sum_{\mu}(-1)^{I_i-M_{I_i}+\mu+1}
\ C(I_d\ I_i\ L;M_{I_d}\ -M_{I_i}\ \mu)\langle N^* I_d ||M_L||N I_i\rangle \nonumber \\*
&&\times
\langle n_d \kappa_d m_d|r^{-(L+1)}\vec{\alpha}\cdot\vec{Y}^{-\mu}_{LL}(\theta,\varphi)|\varepsilon\kappa m\rangle
\ , \nonumber
\end{eqnarray}
where the electronic matrix element can be evaluated in a similar way as
the ones in Eq.~(\ref{Her_me}) to yield
\begin{eqnarray}
&&\langle n_d \kappa_d m_d|
r^{-(L+1)}\vec{\alpha}\cdot\vec{Y}^{-\mu}_{LL}(\theta,\varphi)
|\varepsilon\kappa m\rangle=\\*
&& i(-1)^{j-L+\frac{1}{2}}\sqrt{\frac{(2j+1)(2L+1)}{4\pi L(L+1)}}
C(j\ L\ j_d;m\ -\mu\ m_d)(\kappa_d+\kappa)
\left(\begin{array}{ccc} j_d&j&L\\\frac{1}{2}&-\frac{1}{2}&0\end{array}\right)
\nonumber \\*
&&\times \int_0^{\infty}drr^{-L+1}
\left(g_{n_d \kappa_d}(r)f_{\varepsilon\kappa}(r)
+f_{n_d \kappa_d}(r)g_{\varepsilon\kappa}(r)\right)\ . \nonumber
\end{eqnarray}
This way of writing the electronic matrix element is equivalent to the
more lengthy one presented previously in~\cite{us}.

Now let us consider the matrix element
corresponding to RR. It has, up to the presence of the spherical Bessel
functions, a similar expression,
\begin{equation}
\langle n_d\kappa_d m_d,(m) kLM|H_{er}|\varepsilon\kappa m,0\rangle=
-\sqrt{\frac{4\pi ck}{R}}
\langle n_d\kappa_d m_d|j_L(kr) \vec{\alpha}\cdot\vec{Y}^M_{LL}(\theta,\varphi) |\varepsilon\kappa m\rangle\ .
\end{equation}
Using the properties of the spherical tensor operators~\cite{Grant},
we can write the RR matrix element as
\begin{eqnarray}\label{mger} 
&&\langle n_d\kappa_d m_d,(m) kLM|H_{er}|\varepsilon\kappa m,0\rangle=\\*
&& \sqrt{\frac{4\pi ck}{R}}
i(-1)^{j-L-\frac{1}{2}}\sqrt{\frac{(2j+1)(2L+1)}{4\pi L(L+1)}}C(j\ L\ j_d;m\ M\ m_d)(\kappa_d+\kappa)
\nonumber\\*
&&\times
\left(\begin{array}{ccc} j_d&j&L\\\frac{1}{2}&-\frac{1}{2}&0\end{array}\right)
\int_0^{\infty}drj_L(kr)\left(g_{n_d\kappa_d}(r)f_{\varepsilon\kappa}(r)
+f_{n_d\kappa_d}(r)g_{\varepsilon\kappa}(r)\right)\ . \nonumber
\end{eqnarray}
The remaining matrix element involved in the expression of the Fano
profile parameter $Q_f$ is that of the interaction between the nucleus and the
radiation field~(\ref{eq:hnr}). We make use again of the long-wavelength
approximation, so that the spherical Bessel functions are written as in
Eq.~(\ref{lwl}). With this approximation and using the properties of the
vector spherical harmonics, the magnetic solution of the wave equation
can be  expressed as
\begin{equation}
\vec{A}_{(m)kLM}(\vec{r})=
\sqrt{\frac{4\pi ck}{R}}\frac{k^L}{i\sqrt{L(L+1)}}
\frac{1}{(2L+1)!!}(\vec{r}\times\vec{\nabla})(r^LY_{LM}(\theta,\varphi))\ .
\end{equation}
Rewriting the Hamiltonian $H_{nr}$ we obtain
\begin{equation}
H_{nr}=i\sqrt{\frac{4\pi ck}{R}}\sqrt{\frac{L+1}{L}}\frac{k^L}{(2L+1)!!}\frac{1}{c(L+1)}
\int d^3r_n
(\vec{r}_n\times\vec{j}_n(\vec{r}_n))\cdot\vec{\nabla}(r_n^LY_{LM}(\theta_n,\varphi_n))\ .
\end{equation}
The integral over the nuclear coordinate can be related to the magnetic
multipole operator $M_{LM}$, defined as \cite{Ring}
\begin{equation}
M_{LM}=\frac{1}{c(L+1)}
\int d^3r_n
(\vec{r}_n\times\vec{j}_n(\vec{r}_n))\cdot\vec{\nabla}(r_n^LY_{LM}(\theta_n,\varphi_n))\ .
\end{equation}
The matrix element of the interaction Hamiltonian between the radiation
field and the nucleus yields
\begin{eqnarray}\label{mgnr}
&&\langle N I_i M_{I_i},(m)kLM |H_{nr}|N^* I_d M_{I_d},0\rangle \\*
&&=i\sqrt{\frac{4\pi ck}{R}}\frac{k^L}{\sqrt{L}}\frac{\sqrt{L+1}}{(2L+1)!!}
\langle NI_iM_{I_i}|M_{LM}|N^* I_d M_{I_d}\rangle \nonumber \\*
&&=(-1)^{I_d-M_{I_d}}i\sqrt{\frac{4\pi ck}{R}}\frac{k^L}{(2L+1)!!}\sqrt{\frac{{L+1}}{L(2L+1)}}\nonumber \\*
&&\times C\left(I_d\ I_i\ L;M_{I_d}\ -M_{I_i}\ -M\right)
\langle NI_i\|M_L\|N^* I_d \rangle\ . \nonumber
\end{eqnarray}
Combining the results from Eqs.~(\ref{hmg}), (\ref{mger}) and
(\ref{mgnr}) we write the expression of the dimensionless Fano profile
parameter $Q_f^{(m)}$, making use of the summation properties of the
Clebsch-Gordan coefficients:
\begin{eqnarray}
\frac{1}{Q_f^{(m)}}&=& \frac{\pi\rho_i (-1)^{I_i+3I_d+1}(2j_d+1)}{L(2L+1)(L+1)}
k^{-L}(2L+1)!! \\*
&\times& \sum_{\kappa }(2j+1)(n_d\kappa_d+\kappa)^2
\int_0^{\infty}drr^{-L+1}
\left(g_{n_d\kappa_d}(r)f_{\varepsilon\kappa}(r)
+f_{n_d\kappa_d}(r)g_{\varepsilon\kappa}(r)\right)
\left(\begin{array}{ccc} j_d&j&L\\\frac{1}{2}&-\frac{1}{2}&0\end{array}
\right)^2 \nonumber \\*
&\times& \int_0^{\infty}drj_L(kr)
\left(g_{n_d\kappa_d}(r)f_{\varepsilon\kappa}(r)
+f_{n_d\kappa_d}(r)g_{\varepsilon\kappa}(r)\right) \ . \nonumber
\end{eqnarray}


\section{\label{results} Numerical results}


We have calculated the Fano profile parameter and the interference cross 
section term $\sigma_{\rm int}$ as a function of the incoming electron 
energy for several collisions systems involving electric $E2$ and 
magnetic $M1$ transitions. We consider suitable cases of 
isotopes which have energetically low-lying nuclear levels which make 
the interference between NEEC and RR possible.

For the case of the electric  transitions we consider the
$0^+\to 2^+$ $E2$ transitions of the $^{236}_{92}\mathrm{U}$,
$^{238}_{92}\mathrm{U}$, $^{248}_{96}\mathrm{Cm}$,
$^{174}_{70}\mathrm{Yb}$, $^{170}_{68}\mathrm{Er}$,
$^{154}_{64}\mathrm{Gd}$, $^{156}_{64}\mathrm{Gd}$,
$^{162}_{66}\mathrm{Dy}$ and $^{164}_{66}\mathrm{Dy}$ even-even nuclei. 
The energies of the excited nuclear levels $E_{\rm exc}$
as well as the reduced transition probabilities $B(E2)$, that are
needed for the calculation of the natural width of the nuclear excited
state and the NEEC cross section and rate, are taken from Ref.~\cite{Raman}.
The natural width of the nuclear excited state is considered to be the
sum of the partial radiative rates $A_r^{d \to f}$ and the IC rates
$A^d_{\rm IC}$,
\begin{equation}
\label{width}
\Gamma_d=\sum_f A_r^{d \to f} + \sum_i A^{d \to i}_{\rm IC} \, .
\end{equation}
Here we sum the radiative transition rates to all possible final states 
(note that in our case there is only one nuclear final state, namely, 
the ground state). By summing over $i$ we account for internal conversion 
to the initial state of the NEEC process and all other possible IC 
channels, for the case when the capture occurs into a He-like ion. The IC 
rate can be related to the NEEC rate through the principle of detailed 
balance,
\begin{equation}
A^{d \to i}_{\rm IC}=\frac{2(2I_i+1)}{(2I_d+1)(2j_d+1)}Y_n^{i \to d} \ .
\end{equation} 
The NEEC rates and cross sections are calculated using an improved
version of the computer routines applied in~\cite{us}.
We consider the capture into the bare ions of $^{164}_{66}\mathrm{Dy}$,
$^{170}_{68}\mathrm{Er}$, $^{174}_{70}\mathrm{Yb}$ and
$^{154}_{64}\mathrm{Gd}$. For the cases of the $\mathrm{U}$ isotopes and
for $^{248}_{96}\mathrm{Cm}$, the capture into the $K$ shell is not
possible due to the low energy level of the first excited nuclear state.
For these three systems, recombination into the $L$ shell of initially
He-like ions is the most probable one. We regard the capture of the
electron into a closed shell configuration as a one-electron problem,
without the participation of the $K$-shell electrons. We also consider
the capture of the electron into the He-like ions of
$^{156}_{64}\mathrm{Gd}$ and $^{162}_{66}\mathrm{Dy}$, in which case the
width of the nuclear excited state in Eq.~(\ref{width}) contains 
partial IC rates accounting for the possible IC of the $K$-shell
electrons. 

A numerical evaluation of the radial integrals corresponding to NEEC
[$R_{L,\kappa_d,\kappa}$, see Eq.~(\ref{rrs})] and the ones corresponding
to RR [$I^{\pm}_{L\pm 1}$, Eq.~(\ref{eq:iint})] is needed for the
calculation of the Fano profile
parameters and for the interference cross sections. We consider
Coulomb-Dirac wave functions for the continuum electron and wave
functions calculated with the GRASP92 package~\cite{Par96} by considering a
homogeneously charged nucleus for the bound electron. In the case of
recombination into the He-like ions we assume a total screening of
the nuclear charge for the continuum electron, i.e., we use Coulomb-Dirac
functions with an effective nuclear charge $Z_{\rm eff} = Z-2$.
For the bound electron
wave functions, the electron-electron interaction is accounted for in the
Dirac-Fock approximation. The value of $R_{L,\kappa_d,\kappa}$ is not
affected by finite nuclear size effects on the accuracy level of our
calculations. Nevertheless, the finite size of the nucleus has a
sensitive effect on the energy levels of the bound electron. The energy
of the bound electronic state is calculated with GRASP92 and includes
one-loop one-electron quantum electrodynamic terms, and in the case of
many-electron bound states approximate QED screening corrections. The
nuclear radius $R_0$ is calculated according to the semi-empirical
formula~\cite{Soff}
\begin{equation}
R_0=(1.0793A^{1/3}+0.73587)\,\mathrm{fm}\ ,
\end{equation}
where $A$ is the atomic mass number. Values of the Fano profile
parameters, as well as the NEEC rate and natural width of the nuclear
excited state are presented in Table \ref{Etable}. The values of the
resonance strength of NEEC, given in \cite{us},
\begin{equation}
S_d=\frac{2\pi^2}{p^2}
\frac{A_r^{d \to f} Y_n^{i \to d}}{\Gamma_d} \,,
\end{equation}
are also presented.

\begin{table*}[htb]
\caption{\label{Etable} Parameters of the NEEC total cross section and
the interference term for various heavy ion collision systems involving
electric quadrupole transitions. $E_{\rm exc}$ denotes the nuclear
excitation energy, $E_c$ is the continuum electron energy at resonance,
$Y_{n}$ stands for the resonant recombination rate, and $\Gamma_d$ is the
total width of the excited nuclear state. The column denoted by $S$ contains
the NEEC resonance strengths, $1/Q_f$ is the inverse Fano line profile
parameter, and $R^{\rm int}$ stands for the profile asymmetry parameter.
See the text for further explanations.}
\begin{ruledtabular}
\begin{tabular}{lrrccccrr}
Isotope & $E_{\rm exc}$(keV) & $E_{c}$(keV)  & Orbital &$Y_{n}(1/s)$&$\Gamma_{d}$(eV)& $S$(b eV)&$1/Q_f$ & $R^{\rm int}$ \\ 

\hline
$^{164}_{66}\mathrm{Dy}$ & 73.392 &10.318  & $1s_{1/2}$ &$1.86\times10^8$ &$4.37\times 10^{-8}$   &$3.88	\times10^{-2}$&-2.11$\times10^{-3}$ & 3.67$\times 10^{-3}$\\

$^{170}_{68}\mathrm{Er}$ & 78.591 &11.350  & $1s_{1/2}$ &$2.23\times10^8$ & $5.75\times 10^{-8}$  &$4.70	\times10^{-2}$&-2.07$\times10^{-3}$& 4.05$\times 10^{-3}$\\

$^{174}_{70}\mathrm{Yb}$ & 76.471 &4.897  & $1s_{1/2}$ &$1.79\times10^8$ &$4.85\times 10^{-8}$   &$9.27	\times10^{-2}$&-2.09$\times10^{-3}$&4.30$\times 10^{-3}$\\

$^{154}_{64}\mathrm{Gd}$ & 123.071 &64.005  & $1s_{1/2}$ &$5.69\times10^8$ &$2.51\times 10^{-7}$&$2.91\times10^{-2}$&-2.61$\times10^{-4}$&8.77$\times 10^{-4}$\\

$^{156}_{64}\mathrm{Gd}$ & 88.966 &74.742  & $2s_{1/2}$ &$3.35\times10^7$ & $1.21\times 10^{-7}$  &$7.09\times10^{-4}$&-6.10$\times10^{-5}$&1.67$\times 10^{-3}$ \\
$^{156}_{64}\mathrm{Gd}$ & 88.966&74.896  & $2p_{1/2}$ & $1.16\times10^8$& $1.32\times 10^{-7}$  &$2.25\times10^{-3}$& -1.16$\times10^{-5}$&1.00$\times 10^{-4}$ \\
$^{156}_{64}\mathrm{Gd}$ & 88.966&75.680  & $2p_{3/2}$ & $1.59\times10^8$& $1.27\times 10^{-7}$ &$3.17\times10^{-3}$&3.06$\times10^{-4}$&1.86$\times 10^{-3}$ \\

$^{162}_{66}\mathrm{Dy}$ & 80.660 &65.432  & $2s_{1/2}$ &$2.81\times10^7$ & $9.39\times 10^{-8}$  &$6.25	\times10^{-4}$ &-1.28$\times10^{-4}$ &3.26$\times 10^{-3}$\\
$^{162}_{66}\mathrm{Dy}$ & 80.660 &66.594  & $2p_{1/2}$ & $1.59\times10^8$& $1.11\times 10^{-7}$  &$2.98\times10^{-3}$&-5.78$\times10^{-5}$&3.06$\times 10^{-4}$\\
$^{162}_{66}\mathrm{Dy}$ & 80.660 &66.492  & $2p_{3/2}$ & $2.15\times10^8$& $1.04	\times 10^{-7}$ &$4.24\times10^{-2}$& 3.56$\times10^{-4}$&1.31$\times 10^{-3}$\\

$^{236}_{92}\mathrm{U}$ & 45.242 &12.404  & $2s_{1/2}$ &$1.06\times10^8$ & $1.76	\times 10^{-8}$  &$8.47\times10^{-3}$&1.60$\times10^{-3}$ &2.00$\times 10^{-3}$\\
$^{236}_{92}\mathrm{U}$ & 45.242 &12.698  & $2p_{1/2}$ & $3.02\times10^{9}$& $4.01\times 10^{-7}$  &$1.02\times10^{-2}$&-1.26$\times10^{-3}$&1.27$\times 10^{-3}$\\
$^{236}_{92}\mathrm{U}$ & 45.242 &16.871  & $2p_{3/2}$ & $3.10\times10^{9}$& $2.07\times 10^{-7}$ &$1.52\times10^{-2}$&-9.86$\times10^{-4}$&5.01$\times10^{-4}$\\

$^{238}_{92}\mathrm{U}$ & 44.910 & 12.073    & $2s_{1/2}$ &$1.11\times10^8$ &$1.81\times 10^{-8}$
& $8.80\times10^{-3}$&1.61$\times10^{-3}$&2.01$\times 10^{-3}$\\
$^{238}_{92}\mathrm{U}$ & 44.910 & 12.356  & $2p_{1/2}$ &$3.14\times10^9$ &$4.17\times 10^{-7}$
&$1.06\times10^{-2}$&-1.24$\times10^{-3}$&1.25$\times10^{-3}$\\
$^{238}_{92}\mathrm{U}$ & 44.910 & 16.534 & $2p_{3/2}$ & $3.23\times10^9$ &$2.16\times 10^{-7}$
  &$1.56\times10^{-2}$ &-9.97$\times10^{-4}$&5.07$\times10^{-4}$\\
 
$^{248}_{96}\mathrm{Cm}$ & 43.380 &6.888  & $2s_{1/2}$ &$2.18\times10^8$  &$3.25\times 10^{-8}$&$1.78\times10^{-2}$&1.92$\times10^{-3}$&2.16$\times10^{-3}$\\
$^{248}_{96}\mathrm{Cm}$ & 43.380 &7.190  & $2p_{1/2}$ & $5.47\times10^9$&$7.24\times 10^{-7}$
  &$1.91\times10^{-2}$&-5.96$\times10^{-4}$&5.99$\times10^{-4}$\\
$^{248}_{96}\mathrm{Cm}$ & 43.380 & 12.356  & $2p_{3/2}$ & $5.33\times10^9$&$3.54\times 10^{-7}$&$2.20\times10^{-2}$&-1.43$\times10^{-3}$&7.24$\times10^{-4}$\\
\end{tabular}

\end{ruledtabular}
\end{table*}

The Fano line profile parameter characterizes the strength of the
interference effects between the two recombination channels. Smaller
values of $|Q_f|$ indicate more pronounced interference. A more
quantitative measure of the interference is defined in
Ref.~\cite{rintpaper} as the ratio of the interference term and the resonant
process term at the energy $\varepsilon_{\pm 1/2}=E_d \pm \Gamma_d/2$,
\begin{equation}
R^{\rm int}=\left|\frac{\sigma_{\rm int}(\varepsilon_{\pm 1/2})}{\sigma_{\rm NEEC}(\varepsilon_{\pm 1/2})}\right|=\frac{\Gamma_d}{Y^{i \to d}_n}\frac{2I_d+1}{2I_i+1}\frac{1}{|Q_f|}\ .
\end{equation}
Values for this line asymmetry parameter $R^{\rm int}$ are given in the
last column of Tables~\ref{Etable} and \ref{Mtable}.

A possibility to cross-check the numerical accuracy of the present
calculations is given by the matrix element of the interaction
Hamiltonian $H_{er}$, which enters the expression of the Fano profile
parameter. We can use the matrix element to calculate the total cross
section for RR for a given energy, which can be written in the spherical
wave approach as
\begin{equation}
\sigma_{\rm RR}=\frac{2\pi}{F_i}\frac{1}{2}\sum_{m_s}\frac{1}{4\pi}\int d\Omega_p\sum_{m_d}\sum_{\lambda LM}|\langle n_d\kappa_d m_d,\lambda kLM|H_{er}|\vec{p}m_s,0\rangle|^2\rho_f\ .
\end{equation}
RR cross sections calculated by this formula and with the radial wave
functions described above reproduce the values tabulates in
Ref.~\cite{Eich} with a typical relative accuracy of about one per thousand,
as it can be seen in Table~\ref{tab:rrcs}.
 
\begin{table*}[htb]
\caption{\label{tab:rrtable}
Total RR cross sections for recombination into a given bound state of a
bare ion, compared with results from Ref.~\cite{Eich}. The nuclear
excitation energy $E_{\rm exc}$ is given in the second column. The
values from Ref.~\cite{Eich} are numerically interpolated by a spline
routine to obtain the RR cross section at the resonance energy $E_c$. 
}
\begin{ruledtabular}
\begin{tabular}{lrrccc}\label{tab:rrcs}
 & & & & \multicolumn{2}
 {c}{$\sigma_{RR}$(b)}\\ \cline{5-6}
Isotope & $E_{\rm exc}$(keV) & $E_{c}$(keV) & Orbital &This work&Ref.~\cite{Eich}\\ 
     
\hline

$^{164}_{66}\mathrm{Dy}$ & 73.392 &10.318& $1s_{1/2}$& 832& 832 \\

$^{170}_{68}\mathrm{Er}$ & 78.591 &11.350& $1s_{1/2}$& 797&795\\

$^{174}_{70}\mathrm{Yb}$ & 76.471 &4.897& $1s_{1/2}$ &2080&2080\\

$^{154}_{64}\mathrm{Gd}$ & 123.071 &64.005 & $1s_{1/2}$ &79&79\\

\hline

$^{236}_{92}\mathrm{U}$ & 45.242 &11.113 & $2s_{1/2}$ &245&245 \\
$^{236}_{92}\mathrm{U}$ & 45.242 &11.038 & $2p_{1/2}$ & 295& 294\\
$^{236}_{92}\mathrm{U}$ & 45.242 &15.601 & $2p_{3/2}$ &229&229 \\

$^{238}_{92}\mathrm{U}$ & 44.910 & 10.782 & $2s_{1/2}$ &252&253\\
$^{238}_{92}\mathrm{U}$ & 44.910 &10.706 & $2p_{1/2}$ &306&306\\
$^{238}_{92}\mathrm{U}$ & 44.910 &15.269 & $2p_{3/2}$ & 236&236 \\

$^{248}_{96}\mathrm{Cm}$ & 43.380 &5.500 & $2s_{1/2}$ &543&544\\
$^{248}_{96}\mathrm{Cm}$ & 43.380 &5.398 & $2p_{1/2}$ &768&769 \\
$^{248}_{96}\mathrm{Cm}$ & 43.380 & 11.018 & $2p_{3/2}$ &410& 410\\

\hline

$^{165}_{67}\mathrm{Ho}$ &94.700 & 29.563 & $1s_{1/2}$ &252& 252\\

$^{173}_{70}\mathrm{Yb}$ & 78.647 & 7.073 & $1s_{1/2}$ &1410& 1410\\

$^{185}_{75}\mathrm{Re}$ & 125.358 & 42.198 &$1s_{1/2}$ &212& 212\\

$^{187}_{75}\mathrm{Re}$ & 134.243 &51.083  & $1s_{1/2}$ &166&166 \\

$^{55}_{25}\mathrm{Mn}$ &125.949 & 117.378 & $1s_{1/2}$&0.865&0.849 \\

$^{57}_{26}\mathrm{Fe}$ & 14.412 & 5.135 & $1s_{1/2}$ &216&216\\

$^{40}_{19}\mathrm{K}$ & 29.829 & 24.896 & $1s_{1/2}$ &6.64&6.55 \\

\end{tabular}
\end{ruledtabular}
\end{table*}

For the magnetic multipole transitions we consider the $M1$ transitions
of the odd isotopes $\ ^{165}_{67}\mathrm{Ho}$,
$^{173}_{70}\mathrm{Yb}$, $^{55}_{25}\mathrm{Mn}$,
$^{57}_{26}\mathrm{Fe}$, $^{40}_{19}\mathrm{K}$,
$^{155}_{64}\mathrm{Gd}$, $^{157}_{64}\mathrm{Gd}$,
$^{185}_{75}\mathrm{Re}$ and $^{187}_{75}\mathrm{Re}$. Numerical results
for these ions are presented in Table \ref{Mtable}. We present NEEC
rates and resonance strengths with  improved
accuracy with respect to our previous results~\cite{us}. The electronic
radial integrals are calculated numerically using the same type of wave
functions for the bound and continuum electron as for the electric
transitions. The reduced magnetic transition probability $B(M1)$ and the
energies of the nuclear levels are taken from
Refs.~\cite{NDS1,NDS2,NDS3,NDS4,NDS5,NDS6,NDS7,NDS8,NDS9}. Recombination
into the $K$ shell is possible for all the chosen ions, except for
$\mathrm{Gd}$. We also present results for recombination into the
initially He-like ions of the  $^{155}_{64}\mathrm{Gd}$ and
$^{157}_{64}\mathrm{Gd}$ isotopes.

\begin{table*}[htb]
\caption{\label{Mtable} Parameters of the NEEC total cross section and
the interference term for various heavy ion collision systems involving
magnetic dipole transitions. The notations are as defined in
Table~\ref{Etable}.}
\begin{ruledtabular}
\begin{tabular}{lrrcclccc}
Isotope & $E_{\mathrm{exc}}$(keV) & $E_{c}$(keV) & Orbital &$Y_{n}(1/s)$&$\Gamma_{d}$(eV)& $S$(b eV)&$1/Q_f$&$R^{\rm int}$ \\ 
\hline

$^{165}_{67}\mathrm{Ho}$ &94.700 & 29.563  & $1s_{1/2}$ &1.28$\times 10^{10}$&1.17$\times 10^{-5}$ &8.84$\times 10^{-1}$ &-1.67$\times10^{-3}$&2.90$\times10^{-3}$\\

$^{173}_{70}\mathrm{Yb}$ & 78.647 & 7.073  & $1s_{1/2}$ &7.32$\times10^9$ &4.80$\times 10^{-6}$ & 1.26&-2.24$\times10^{-3}$&2.98$\times10^{-3}$\\

$^{185}_{75}\mathrm{Re}$ & 125.358 & 42.198  & $1s_{1/2}$ &2.62$\times 10^{10}$ &2.36$\times 10^{-5}$  & 1.34 &-2.58$\times10^{-3}$&4.71$\times10^{-3}$\\

$^{187}_{75}\mathrm{Re}$ & 134.243 &51.083   & $1s_{1/2}$ & 2.50$\times 10^{10}$& 2.47$\times 10^{-5}$& 1.16&-2.50$\times10^{-3}$&5.00$\times10^{-3}$\\

$\ ^{55}_{25}\mathrm{Mn}$ &125.949 & 117.378  & $1s_{1/2}$ &2.45$\times10^7$ &1.75$\times 10^{-6}$& 9.22$\times10^{-4}$&-2.14$\times10^{-5}$&3.10$\times10^{-3}$\\

$^{57}_{26}\mathrm{Fe}$ & 14.412 & 5.135  & $1s_{1/2}$ &6.21$\times10^6$ & 2.56$\times 10^{-9}$& 1.19$\times10^{-3}$&-6.73$\times10^{-5}$&8.42$\times10^{-5}$\\

$^{40}_{19}\mathrm{K}$ & 29.829 & 24.896  & $1s_{1/2}$ &1.33$\times10^7$& 9.47$\times 10^{-8}$& 2.27$\times10^{-3}$&-1.46$\times10^{-5}$&1.22$\times10^{-4}$\\
 
$^{155}_{64}\mathrm{Gd}$ & 60.008 &45.784   & $2s_{1/2}$ &2.73$\times10^8$  &
1.97$\times10^{-6}$&3.18$\times10^{-3}$&-1.25$\times10^{-4}$&2.06$\times10^{-3}$\\
$^{155}_{64}\mathrm{Gd}$ & 60.008 &45.938  & $2p_{1/2}$ &2.40$\times10^7$  &
1.86$\times10^{-6}$&2.94$\times10^{-4}$&-1.85$\times10^{-5}$&3.27$\times10^{-3}$\\
$^{155}_{64}\mathrm{Gd}$ & 60.008 &46.722  & $2p_{3/2}$ & 4.00$\times10^6$&
1.85$\times10^{-6}$&4.84$\times10^{-5}$&-1.81$\times10^{-5}$&1.91$\times10^{-2}$\\

$^{157}_{64}\mathrm{Gd}$ & 54.533 &40.309   & $2s_{1/2}$ &4.16$\times10^8$ & 4.37$\times 10^{-7}$& 2.86$\times10^{-2}$&-1.25$\times10^{-4}$&3.00$\times10^{-4}$\\
$^{157}_{64}\mathrm{Gd}$ & 54.533 &40.463   & $2p_{1/2}$ & 3.68$\times10^7$ &2.71$
\times10^{-7}$& 4.07$\times10^{-3}$&-2.00$\times10^{-5}$&3.36$\times10^{-4}$\\

$^{157}_{64}\mathrm{Gd}$ & 54.533 &41.247   & $2p_{3/2}$ & 6.21$\times10^6$ &
2.56$\times10^{-7}$& 7.12$\times10^{-4}$&-1.94$\times10^{-5}$&1.82$\times10^{-3}$

\end{tabular}
\end{ruledtabular}
\end{table*}

\begin{figure}
\begin{center}
\includegraphics[width=0.60\textwidth]{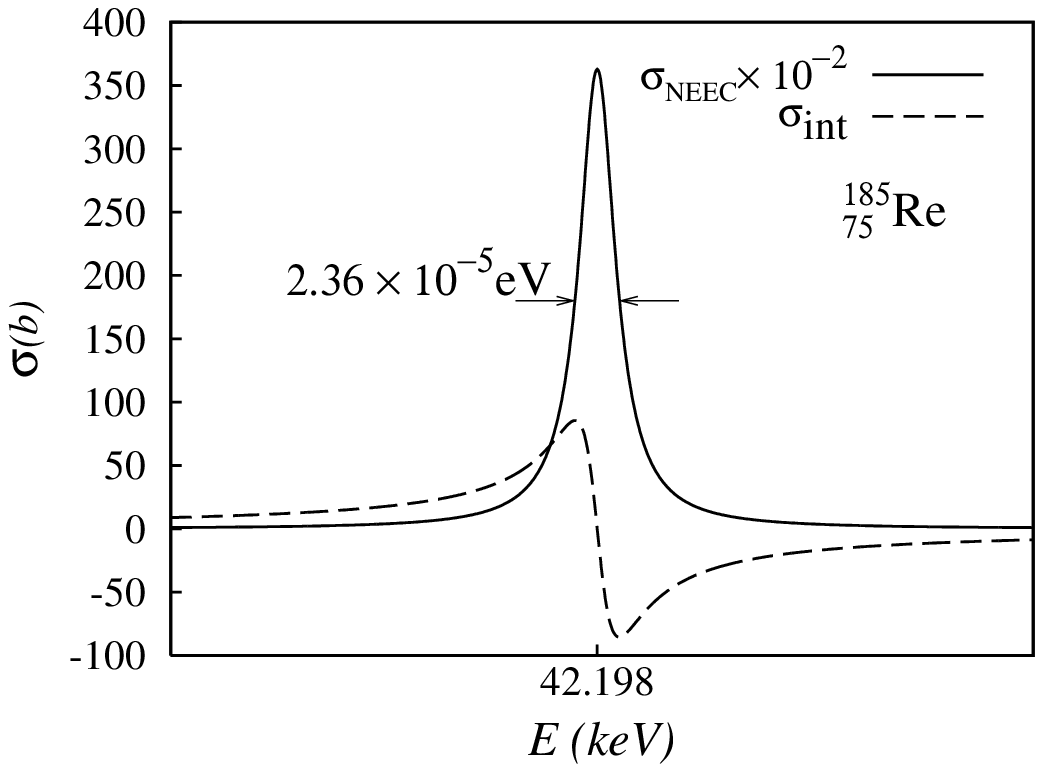}
\includegraphics[width=0.60\textwidth]{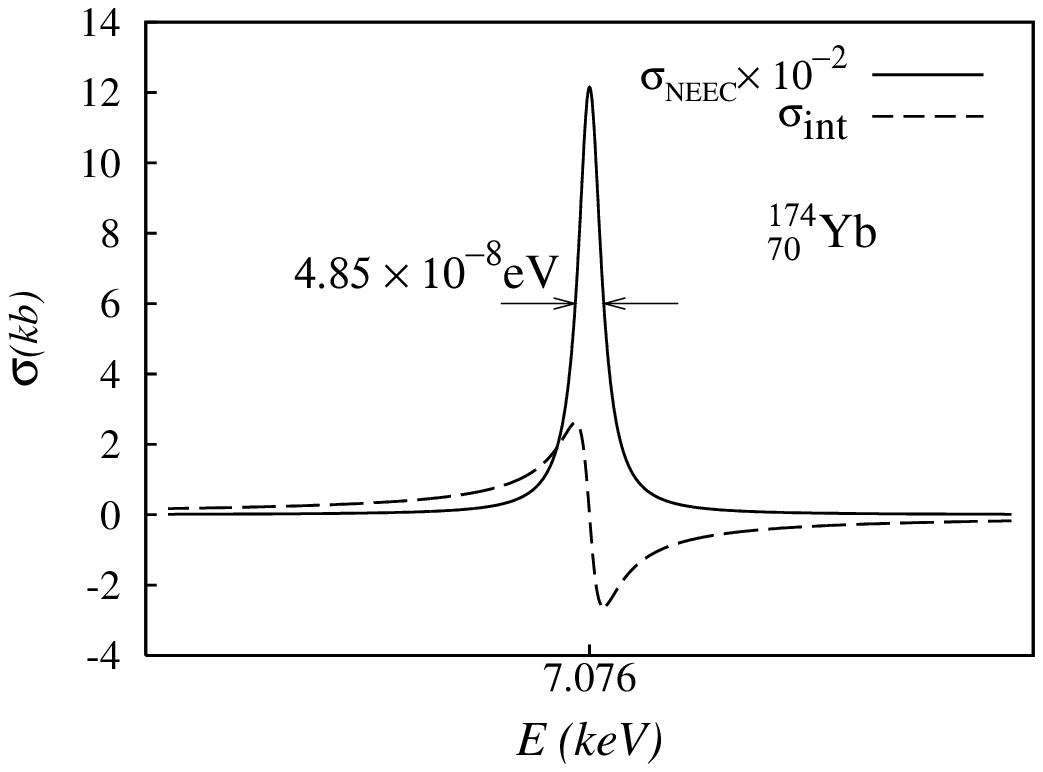}
\caption{\label{yb}Interference and NEEC terms of the cross section for
capture into bare $^{185}_{75}$Re (upper figure) and bare $^{174}_{70}$Yb
(lower figure) ions as a function of the continuum electron energy. The NEEC
term is scaled by a factor of $10^{-2}$. }
\label{plots}
\end{center}
\end{figure}

In Fig.~\ref{yb} interference and scaled NEEC cross section terms are
plotted as a function of the continuum electron energy for the $M1$
transition of $^{185}_{75}\mathrm{Re}$ and $E2$ transition of
$^{174}_{70}\mathrm{Yb}$, respectively. These are the isotopes with the
largest values for the resonance strengths for the magnetic and electric
multipole transitions, respectively. The NEEC cross section has the
shape of a very narrow Lorentzian, with the width given by the natural
width of the excited nuclear state, about 2.4$\times 10^{-5}$~eV for the
case of $^{185}_{75}\mathrm{Re}$ and 4.9$\times 10^{-8}$~eV for the case
of $^{174}_{70}\mathrm{Yb}$. The interference term $\sigma_{\rm int}$
for both electric and magnetic cases is more than two orders of
magnitude smaller than the NEEC terms $\sigma_{\rm NEEC}$. The magnitude
of the interference term can be explained by investigating the
contributions of the multipolarities that enter in the RR cross section
$\sigma_{\rm RR}$.  While $\sigma_{\rm RR}$ consists of an infinite sum
of multipolarities, in the interference process only the RR photon with
the multipolarity of the nuclear transition participates.  The
main contribution to the RR cross section comes from the electric dipole
$E1$ photon. The cross sections corresponding to the $M1$ and $E2$
photons are considerably smaller. In the case of
$^{174}_{70}\mathrm{Yb}$, the $E2$ multipole accounts for only 121~b
in the total RR cross section of 2080~b, while the $M1$ multipole for
$^{185}_{75}\mathrm{Re}$ only contributes 0.5~b to the total RR cross
section of 212~b. 

\begin{figure}
\begin{center}
\includegraphics[width=0.60\textwidth]{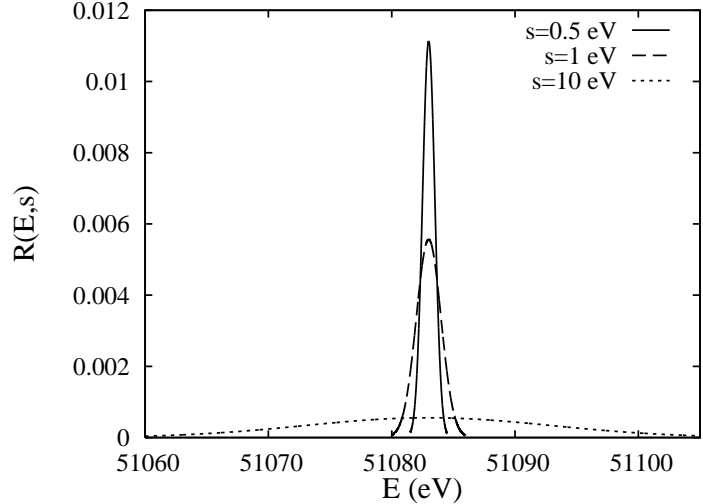}
\end{center}
\caption{The ratio $R(E,s)$ in Eq.~(\ref{r_s}) for recombination into
bare rhenium as a function of the energy of the continuum electron for
three different experimental electron energy width parameters $s$. See
text for further explanations.} 
\label{ratio}
\end{figure}

As an electron energy resolution in the order of $10^{-5}$ eV and less can
not be presently achieved in an experiment, we convolute the theoretical
total cross section with the energy distribution of the electrons to
give an orientation for measurements in  near future.  The energy
distribution of the incoming electrons is assumed to be described by a
Gaussian function with a width parameter~$s$. The RR cross section has
a practically constant value on the energy interval of~$s$. In order to
demonstrate the magnitude of the NEEC and interference cross sections
$\sigma_{\rm NEEC}$ and $\sigma_{\rm int}$ compared to that of RR, we
present in Fig.~\ref{ratio} the ratio of the convoluted cross sections,
\begin{equation}\label{r_s}
R(E,s)=\frac{\tilde{\sigma}_{\rm
NEEC}(E,s)+\tilde{\sigma}_{\rm int}(E,s)}{\tilde{\sigma}_{\rm RR}(E,s)} \,,
\end{equation}
in the case of $^{185}_{75}$Re as a function of the continuum electron
energy for the three different experimental width parameters $s=0.5$~eV,
$1$~eV and $10$~eV. While for a width parameter $s=0.5$~eV the
contributions of the NEEC and interference terms can be clearly
discerned from the RR background, for presently more realistic widths in
the order of  eVs or tens of eV the values of the ratio $R(E,s)$
are too small to be observed experimentally.

\section{\label{sum} Summary}


In this article we investigated the interference between NEEC and RR in
an electron recombination process. We derived the interference cross
section and expressed it with the help  of the dimensionless Fano profile
parameter.

We calculated the interaction matrix elements for both electric and
magnetic multipolarities using relativistic electronic wavefunctions.
Nuclear excitations are described using a phenomenological nuclear
collective model. The nuclear part of the matrix element is written by
the help of the reduced nuclear transition probability whose value is
taken from experimental works. For the quantization of the radiation
field we use the multipole expansion.

Numerical values for the Fano profile parameters and interference cross 
sections were obtained for various heavy-ion collision systems. The 
interference term in the total cross section of the recombination 
process is about two orders of magnitude smaller than the NEEC cross 
section. This is associated with the fact that from the infinite 
multipole expansion of the RR radiation, only the multipolarities 
corresponding to the type of nuclear transition interfere with the 
radiative decay photons following NEEC. The interference term has a 
narrow extent on the electron energy scale, which is related to the 
small natural width of the nuclear excited state. In order to simulate 
data of a recombination experiment, we convolute the total cross section 
with a Gaussian electron energy distribution of realistic width 
parameters. While for well-defined experimental electron energies the 
presence of NEEC could be discerned from the RR background, for larger 
width parameters both NEEC and the interference with RR become difficult 
to be observed experimentally.

If the angular distribution of the emitted photons in the radiative 
decay of the nucleus following NEEC is different from the one of the RR 
photons, this can be used to identify the resonant process in the RR 
background. Calculations investigating a possible NEEC signature in the 
angular distribution of the emitted electrons are in progress.


\begin{acknowledgments}

A. P. acknowledges the support from the Deutsche Forschungsgemeinschaft 
(DFG).

\end{acknowledgments}

\bibliography{interf}

\end{document}